# FAST OUTPUT ENERGY REGULATION IN A MEDICAL PROTON LINAC

L. Yu. Ovchinnikova[1,2,†], A. P. Durkin[1], A. S. Kurilik[1], V. V. Paramonov[1,‡]
[1]INR RAS, Moscow, Russia
[2]Ferrite Domen Co., St. Petersburg, Russia

# БЫСТРАЯ РЕГУЛИРОВКА ЭНЕРГИИ В ЛИНЕЙНОМ УСКОРИТЕЛЕ ПРОТОНОВ ДЛЯ МЕДИЦИНЫ

Л. Ю. Овчинникова[1,2,†], А. П. Дуркин[1], А. С. Курилик[1], В. В. Парамонов[1,‡]
[1]ФГБУН Институт Ядерных Исследований РАН, Москва, Россия
[2]АО «НИИ «Феррит-Домен», Санкт-Петербург, Россия

(6 October 2023)

*Abstract*

In proton therapy, depth scanning of the irradiated object is performed by changing the Output Energy (OE) of the accelerated beam. In pulsed linear accelerators, adjustment of the OE is usually by changing the amplitude and/or phase of the field in the accelerating elements from one RF pulse to another. The application of non-inertial traveling wave accelerating sections makes it possible to change quickly the phase of the accelerating field during the RF pulse. The phase of the field in the constant gradient section is determined both by the phase of the input RF signal and by the process of wave propagation in the dispersive structure. The calculation results of the traveling wave propagation in the accelerating structure when the phase of the input RF signal changes and the results of simulation the dynamics of particles confirm the change in the linac's OE during the RF pulse. The proposed method for regulation the OE makes it possible to increase in orders the speed of scanning the irradiated object by depth.

*Аннотация*

В протонной терапии сканирование объекта облучения по глубине производится изменением Выходной Энергии (ВЭ) ускоренного пучка. В импульсных линейных ускорителях принята регулировка ВЭ за счёт изменения амплитуды и/или фазы поля в ускоряющих элементах от одного СВЧ импульса к другому. Применение являющихся безынерционными ускоряющих секций на бегущей волне позволяет быстрее изменять фазу ускоряющего поля в течение ВЧ импульса. Фаза и амплитуда поля в секции определяется как фазой и амплитудой входного СВЧ сигнала, так и процессом его распространения в дисперсной структуре. Результаты расчёта распространения бегущей волны в ускоряющей структуре при изменении фазы СВЧ сигнала и результаты моделирования динамики частиц, подтверждают изменение ВЭ ускорителя в процессе СВЧ импульса. Предлагаемая методика регулировки ВЭ позволяет многократно повысить скорость сканирования объекта облучения по глубине.

## ВВЕДЕНИЕ

Предложение разрабатываемого импульсного линейного ускорителя (ЛУ) протонов с выходной энергией частиц до 230 МэВ изложено в [1]. Ускоряющими элементами в основной части ЛУ являются секции структуры на бегущей волне (Traveling Wave – TW) - с постоянным градиентом. Регулировка ВЭ ЛУ, необходимая для сканирования объекта по глубине, за счёт изменения амплитуды ускоряющего поля ведёт к необходимости управлять параметрами мощного высоковольтного оборудования. Это возможно от одного СВЧ импульса к другому, но технически не реализовано в течение одного короткого ВЧ импульса длительностью не более десятков мкс.

Изменение фазы ВЧ поля можно производить на низком уровне СВЧ мощности. Моделирование выходного СВЧ импульса клистрона КИУ-286 [2], рассматриваемого для ЛУ, при скачкообразном изменении фазы входного сигнала показало в выходном сигнал отслеживание изменения с незначительным переходным процессом длительностью ~0.3 мкс.

В ускоряющей структуре на стоячей волне изменение амплитуды или фазы поля происходит за время $\sim 3\tau$, где $\tau$ - постоянная времени резонатора. Для структур S частотного диапазона $3\tau \approx 4.5$ мкс, что практически определяет возможность изменения ВЭ только от одного СВЧ импульса к другому.

В отличии от резонаторов, TW структура является безынерционной. Принцип регулировки ВЭ пучка в TW структуре при скачкообразном изменении фазы входного СВЧ импульса сформулирован в работе [3]. Используется возможность иметь в TW структуре на различных участках СВЧ поля с одинаковой амплитудой, но разными фазами колебаний. Граница раздела между участками перемещается вдоль структуры с

† lubovch@inr.ru
‡ paramono@inr.ru

групповой скоростью волны. При этом обеспечивается плавное и быстрое изменение ВЭ.

Характерным масштабом времени процесса регулировки становится время заполнения TW структуры. Для разрабатываемого ЛУ это ~1 мкс. Возможность дискретного изменения фазы входного СВЧ сигнала на небольшую величину на низком уровне СВЧ мощности за такой промежуток времени доказана результатами работы [4].

В данной работе прямым численным моделированием движения протонов в СВЧ поле распространяющейся вдоль TW структуры составной волны подтверждена возможность быстрого изменения ВЭ ЛУ в течение СВЧ импульса и определены особенности и количественные характеристики выходного пучка. Приведены особенности поля дозы облучения при таком способе регулировки ВЭ ЛУ.

Исследование проведено на примере регулировки СВЧ поля в выходной, 12-й TW структуре ЛУ.

## МОДЕЛИРОВАНИЕ РАСПРОСТРАНЕНИЯ СВЧ СИГНАЛА В TW-СТРУКТУРЕ

Методика и основные результаты моделирования распространения СВЧ импульса в дисперсной TW-структуре с постоянным градиентом описана в работе [5]. Рассматриваемая TW структура работает на волне вида $2\pi/3$, содержит 102 ячейки и имеет отрицательную дисперсию [3]. Устройство ввода СВЧ мощности (УВМ) располагается в последней 102-й ячейке и СВЧ волна движется навстречу пучку. Структура рассчитана на ускорение протонов от 200 до 230 МэВ. На вход в УВМ подаётся СВЧ сигнал, фаза которого меняется скачкообразно на 8° каждую 1 мкс, общая длительность моделируемого процесса 7 мкс. Поле в структуре можно описать соотношением:
$$E = E_0 \cos(\omega t + \varphi_r(n,t)), \quad (1)$$
где $E_0$ - амплитуда ускоряющего поля, а $\varphi_r(n,t)$ - изменяющаяся во времени для каждой ячейки с номером $n$ фаза, описывающая распространение волны от УВМ.

На Рис. 1 для трех ячеек иллюстрируется процесс распространения волны в дисперсной TW структуре. Серый фон на рисунках соответствует амплитуде поля $E_0$, а черные линии – эффективной напряжённости поля (1) для выбранных ячеек, обусловленной скачкообразным изменением фазы на входе в УВМ.

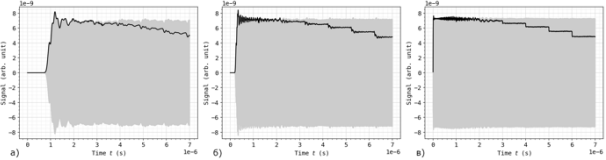

Рисунок 1: Эффективная напряжённость поля в ячейке 1 - вход пучка (а), в ячейке 51 (б) и в ячейке 102 с УВМ (в).

Из приведённых рисунков видно, что первоначально скачкообразный фронт распространения изменения фазы вблизи ячейки с УВМ, Рис. 1(в), расплывается при распространении по структуре, Рис. 1(б). При этом амплитуда поля не меняется. А Рис. 1(а) подтверждает ожидаемое время заполнения структуры – фронт изменения фазы доходит до первой ячейки после первой микросекунды. Это подтверждается и обобщённой двумерной диаграммой на Рис. 2. Для обеспечения постоянства ускоряющего градиента групповая скорость изменяется по длине структуры и в левой части Рис. 2 отчётливо видна нелинейная зависимость времени прихода фронта изменения фазы от номера ячейки.

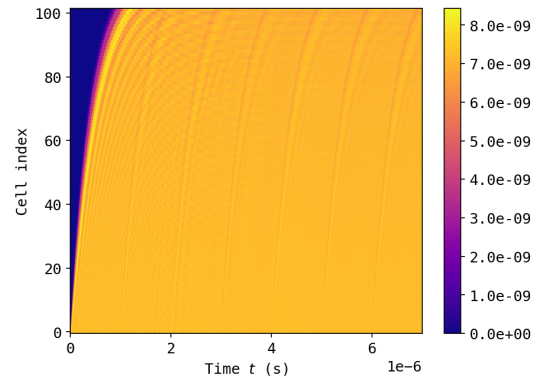

Рисунок 2: Напряжённость поля в ячейках. По горизонтальной оси – время, по вертикальной – номер ячейки. 0 соответствует ячейке с УВМ. Цветом показана амплитуда напряжённости поля.

Приведённые на Рис. 1 и Рис. 2 зависимости полностью соответствуют физическим представлениям о распространении узкополосного сигнала в дисперсной структуре.

## ДИНАМИКА ПРОТОННОГО ПУЧКА В СВЧ ИМПУЛЬСЕ ПРОТОННОГО УСКОРИТЕЛЯ

Движение протонов в ускоряющем СВЧ поле описывается уравнениями (2) где $\lambda$, $\beta_c$ - длина и относительная фазовая скорость волны в ячейке, $W_0$ - энергия покоя протона, $\beta$, $\gamma$ - относительная скорость и Лоренц-фактор частицы, $\varphi_s = -10°$ - расчётное значение синхронной фазы, $\tau = ct$, $\varphi = \omega\tau/c - 2\pi z/\lambda\beta_c$ - нормированное время и фаза частицы.

$$\begin{cases} \dfrac{d\varphi}{dz} = \dfrac{2\pi}{\lambda}\left[\dfrac{1}{\beta} - \dfrac{1}{\beta_c}\right] \\ \dfrac{d\beta}{dz} = \dfrac{eE_0 \cos(\varphi + \varphi_s + \varphi_r(n,t'))}{W_0 \beta \gamma^3} \end{cases} \quad (2)$$

Расчётная величина ускоряющего поля $E_0$ для рассматриваемой структуры равна 14.67 МВ/м. Система (2) решается последовательно от ячейки к ячейке. $\varphi_r(n,t')$, следующая из результатов расчёта распространения волны, однозначно связана с номером ячейки $n$ и временем $t'$ от начала изменения фазы.

Продольные фазовые портреты пучка на входе в структуру [1], и выходе в режиме номинального ускорения, т.е. без изменения фазы СВЧ сигнала, показаны на Рис. 3.

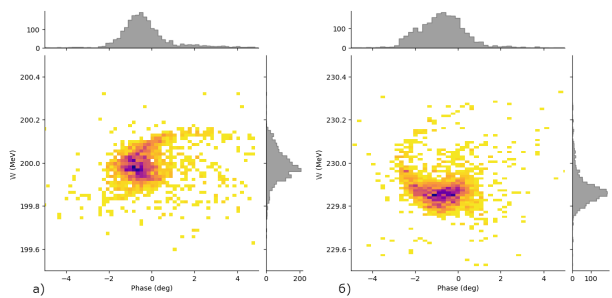

Рисунок 3: Продольные фазовые портреты пучка на входе (а) и выходе (б) структуры в режиме номинального ускорения.

Моделирование регулировки энергии пучка проведено решением во времени методом Рунге-Кутта системы уравнений (2) с использованием полученных ранее распределений поля распространяющейся волны.

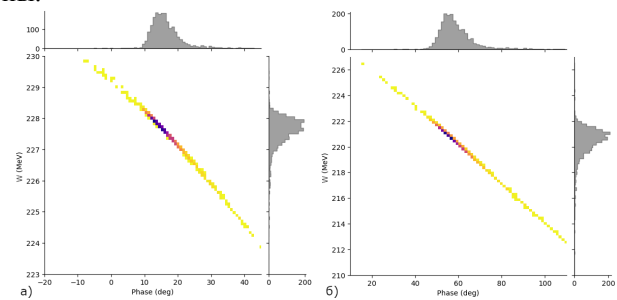

Рисунок 4: Продольные фазовые портреты пучка на выходе структуры, при сдвиге фазы на СВЧ поля на входе УВМ на 30° (а) и 40° (б).

Для существенной регулировки ВЭ необходимо нарушить условия устойчивого продольного движения частиц в структуре, т.е. создать условия, когда принцип автофазировки выполняться не может. При этом продольное движение частиц качественно отличается от номинального режима. На Рис. 4 показаны продольные фазовые портреты частиц на выходе структуры при значительном изменении фазы СВЧ сигнала. Следует обратить внимание на различие шкал этих рисунков. В единой шкале на одном из рисунков распределение выродится в пятно.

При расчётной синхронной фазе структуры -10° для нарушения автофазировки необходимо ввести изменение фазы поля не менее чем ~20°. Это подтверждается приведёнными ниже данными моделирования на Рис. 5 и 6 - до 2 мкс выбранного закона изменения фазы поля на входе в УВМ действует режим автофазировки – при изменении фазы поля на 8° и 16° энергия ускоренного пучка существенно не меняется. Начиная с 2.5 мкс виден выход из области устойчивого продольного движения, что соответствует 16° изменения фазы поля, выходная энергия ускоренного пучка начинает меняться.

Как видно из приведённых на Рис. 4 распределений, при значительном изменении фазы входного сигнала изменяется не только средняя энергия частиц в сгустке, но и значительно увеличивается разброс частиц по энергии.

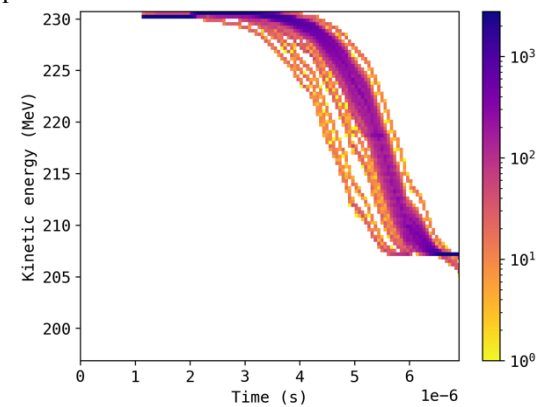

Рисунок 5: Распределение количества частиц на выходе ЛУ по времени и энергии.

При выбранном законе изменения входной фазы распределение числа частиц по времени и энергии показано на Рис. 5.

Графики зависимостей средней энергии частиц в сгустке и среднеквадратичного разброса частиц по энергии для выбранного закона изменения фазы СВЧ поля приведены на Рис. 6.

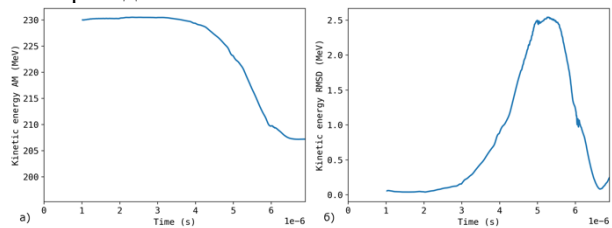

Рисунок. 6: Зависимости от времени средней энергии пучка (а) и среднеквадратичного разброса энергии частиц в сгустке (б).

Полученные результаты моделирования движения протонов в меняющемся в течение СВЧ импульса поле совпадают, и независимо подтверждают, поскольку получены с помощью вновь разработанного специализированного программного обеспечения, полученные ранее в [1] характеристики пучка при изменении фазы поля от одного СВЧ импульса к другому. Но это следует рассматривать как утверждение, что при реализации изменения фазы в течение СВЧ импульса будут получены аналогичные параметры пучка, но на несколько порядков быстрее по времени и с более широкими и гибкими возможностями формирования поля облучения.

## ДОЗОВЫЕ ПОЛЯ, СФОРМИРОВАННЫЕ УСКОРЕННЫМИ СГУСТКАМИ

С использованием Geant4 [6] было выполнено моделирование распределения поглощённой дозы в воде по глубине от серии ускоренных сгустков (рис. 5).

Рассчитанное с применением Geant4 распределение дозы облучения по глубине от серии ускоренных сгустков, соответствующей временному промежутку регулирования ВЭ от 3 до 6 мкс, см. Рис. 5, приведено на Рис. 7. Область облучения расположена на глубине от 280 до 320 мм.

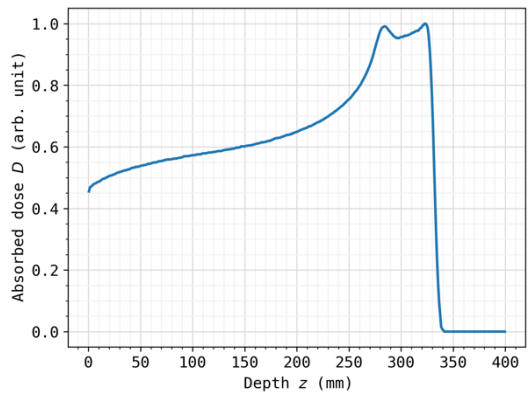

Рис. 7: Распределение поглощённой дозы в воде по глубине.

Существуют различия в формировании дозы при регулировке ВЭ в течение СВЧ импульса от рассмотренного в [7] формирования при регулировке ВЭ от СВЧ импульса к СВЧ импульсу.

За время 1 мкс через структуру пролетает 476 сгустков. При регулировке ВЭ от импульса к импульсу все эти сгустки формируют одну достаточно узкую зону [7], с незначительно размытым пиком Брэгга из-за разброса по энергии частиц в сгустке. При регулировке ВЭ в течение импульса все сгустки различаются по средней энергии и это различие, как можно оценить по графикам на Рис. 5 и 6, меньше или соизмеримо с разбросом по энергии частиц в сгустке. Поэтому размытые разбросом по энергии частиц в сгустке пики Брэгга от соседних сгустков перекрываются. Это позволяет получить более равномерное распределение дозы по глубине в отличие от набора пиков Брэгга, соответствующих отдельным значениям энергии. Набор величины дозы может осуществляться за счёт повторения импульсов.

Неравномерность распределения дозы в области облучения, показанная на Рис. 7 – выбросы по краям области, обусловлена выбором временного промежутка отбора частиц.

Для решения обратной задачи – определения закона изменения фазы в течение СВЧ импульса по заданному в плане облучения распределению дозы, применима изложенная в [8] методика. Обеспечивая изначально более плавное распределение дозы, регулировка ВЭ в течение СВЧ импульса по крайней мере не снижает эффективности применения этой методики.

## ЗАКЛЮЧЕНИЕ

Разработаны и реализованы в вычислительных программах методики моделирования как распространения СВЧ волны в дисперсной TW структуре, так и динамики частиц в поле этой волны. Результаты проведённых расчётов подтверждают возможность регулировки выходной энергии ускорителя в течение СВЧ импульса. По сравнению с регулировкой от импульса к импульсу, при аналогичных параметрах пучка такая регулировка обеспечивает на порядки более высокую скорость сканирования по глубине при более равномерном распределении дозы облучения. При этом, как минимум, не усложняется задача планирования распределения дозы при облучении.